\documentclass[a4paper,11pt]{article}

\usepackage{jinstpub} 
\usepackage{wasysym}
\usepackage{xspace}
\usepackage{xcolor}
\usepackage{color}
\usepackage{float}
\usepackage[normalem]{ulem}

\usepackage{lineno}
\newcommand{\vanode}{\ensuremath{V_{\textrm{a}}}\xspace}
\newcommand{\vback}{\ensuremath{V_{\textrm{b}}}\xspace}
\newcommand{\vmesh}{\ensuremath{V_{\textrm{m}}}\xspace}
\newcommand{\vpmt}{\ensuremath{V_{\textrm{PMT}}}\xspace}
\newcommand{\dab}{\ensuremath{\Delta V_{\textrm{ab}}}\xspace}
\newcommand{\edrift}{\ensuremath{E_{\textrm{drift}}}\xspace}

\newcommand{\qdrift}{\ensuremath{q_{\textrm{drift}}}\xspace}
\newcommand{\qtot  }{\ensuremath{q_{\textrm{tot}}}\xspace}

\newcommand{\registered}{\ensuremath{^{\textrm{\textregistered}}}\xspace}

\newcommand{\ignoreblock}[1]{}

\title{\boldmath Electroluminescence and charge multiplication in liquid xenon with a VCC-like Microstrip Plate}

\author[a,b,1,2]{G. Mart\'inez-Lema,\note{Corresponding author.}\note{Now at Institut de F\'isica Corpuscular, Spain}}
\author[c]{V. Chepel,}
\author[a]{and A. Breskin}

\affiliation[a]{Dept. of Astrophysics and Particle Physics, Weizmann Institute of Science, Rehovot, Israel}
\affiliation[b]{Unit of Nuclear Engineering, Ben-Gurion University of the Negev, Beer-Sheva, Israel}
\affiliation[c]{LIP-Coimbra and Department of Physics, University of Coimbra, 3004-516 Coimbra, Portugal}

\emailAdd{gonzalo.martinez@uv.es}

\abstract{

We report on the first observation of electroluminescence and charge amplification with a Virtual Cathode Chamber (VCC) microstrips plate immersed in liquid xenon. Both were observed in an intense non-uniform electric field in the vicinity of 2-$\mu$m narrow anode strips deposited, with a 2~mm pitch, on a semiconductive glass substrate (S8900), with a cathode film on its backside. An initial light yield of $\sim$460 VUV photons per drifting electron was measured, which degraded within tens of minutes stabilizing at (27.0~$\pm$~3.1)~photons per electron. The electroluminescence was accompanied by electron multiplication with an estimated charge gain $\lesssim$5. Further investigations are necessary to understand and mitigate the light yield degradation phenomenon. We expect other substrate materials, including VUV-transparent ones, to provide large stable photon yields, compatible with our model predictions. The VCC configuration has demonstrated great potential in single-phase noble-liquid detectors, particularly for dark-matter searches, neutrino physics and other fields.}

\keywords{
\\ Charge transport, multiplication and electroluminescence in rare gases and liquids
\\ Time Projection Chambers
\\ Noble liquid detectors
\\ Micropattern gaseous detectors
}

\begin{document}
\maketitle
\flushbottom

\section{Introduction}
\label{sec:intro}

Noble-liquid detectors have been playing a major role in physics experiments and in other applications, as reviewed in \cite{AprileDoke:2010,Chepel:2013,Akimov:2021book}.
Copious ionization-electron yields, deposited by charged particles’ tracks in massive liquid argon (LAr) neutrino experiments \cite{Majumdar_2021,DUNE:2024wvj}, are effectively detected, without multiplication, by wires or perforated electrodes deployed within the liquid – in the so-called single-phase mode.
The detection of small energy deposits, like that by Dark-matter (DM) particle-induced nuclear recoils in the noble liquid (LAr, LXe) require amplification: charge-avalanche multiplication (CM) or electroluminescence (EL).
Both processes occur in noble liquids under very intense electric fields ($\sim 10^5 - 10^6$~V/cm \cite{MASUDA1979247,PhysRevA.9.2582,DOKE198287,Aprile:2014ELthreshold}).
Such fields were difficult to reach, in robust configurations and stable conditions – despite the major efforts made over the past five decades. For recent attempts with thin-wire configurations see \cite{Aprile:2014ELthreshold,Juyal_2021,Lin_2021,Tönnies_2024}.
Therefore, most DM experiments deploy dual-phase detectors \cite{Akimov:2021book,Dolgoshein_1970}; recoil-induced electrons are extracted from the liquid target into the gas phase, where amplification occurs at orders-of-magnitude
lower electric fields, mostly by EL induced by the extracted electrons over a narrow gas gap.
This became the most powerful class of instruments in DM searches for Weakly Interacting Massive Particles (WIMPs) and for detection of neutrino-nucleus scattering (see \cite{AprileDoke:2010,Chepel:2013,Akimov:2021book} and references therein).

To be sensitive to WIMP-nucleon cross-sections down to the neutrino floor, the target mass of future DM detectors is in the tens of tons scale for LXe \cite{Aalbers_2023} and 20 kton for LAr~\cite{DarkSide-20k:2017zyg}.
Scaling up of current dual-phase detectors may encounter some technical difficulties such as EL-gap variations due to mesh-electrode sagging over large areas, electrical instabilities at the liquid-gas interface (e.g. spontaneous delayed-electron emission) and electron-extraction inefficiencies from liquid-to-gas, which may seriously affect their performance.
Some recent works on novel dual-phase detector concepts, with EL readout, aimed at solving such issues \cite{Erdal:2017yiu, Chepel_2023, Martinez-Lema_2024_etransfer}; others focused on enhanced gain stability in CM mode \cite{TesiRWELL, Tesi_2024}.
A single-phase detector would have numerous advantages which has motivated the search for novel "wire-less" concepts
Besides the issues of inefficient electron transfer, liquid-to-gas interface instabilities and EL gap variations mentioned above, single-phase detectors can be deployed in both vertical and horizontal geometries; the drift volume can be divided into two symmetric sections, each operating at half the considerable drift potential.

The concept presented here is a continuation of the effort of detection of VUV photons resulting from electron-induced EL and CM at the vicinity of narrow thin strips - deposited on an insulating substrate immersed in the noble liquid.
Policarpo  \cite{POLICARPO1995568} validated CM of a factor $\sim$10 with such microstrip plate (MSP) having 8-$\mu$m narrow anode strips deposited on thin glass, interlaced by broader cathode strips; the charge multiplication was limited by discharges occurring between the adjacent anode and
cathode strips.
This so-called Microstrip Gas Chamber (MSGC) was developed originally by Oed \cite{OED1988351} for charged-particle imaging.
Following single-phase detector ideas presented in \cite{Breskin:2022novel}, we have undertaken systematic model-calculations and experimental investigations focusing on optimizing parameters of various MSP configurations with the aim of maximizing their combined EL and CM photon yields.
The results with a MSGC detector of the configuration studied in \cite{POLICARPO1995568} in LXe, are presented in \cite{Martinez-Lema_2024_microstrips}, where photo yields of $\sim$30 photons per electron were attained at the maximal attainable inter-strip potential of 2~kV.
Preliminary investigations (unpublished) have been made with this MSGC plate also in LAr \cite{lidine_2024}.

\begin{figure}[b]
    \centering
    \includegraphics[width=0.49\textwidth]{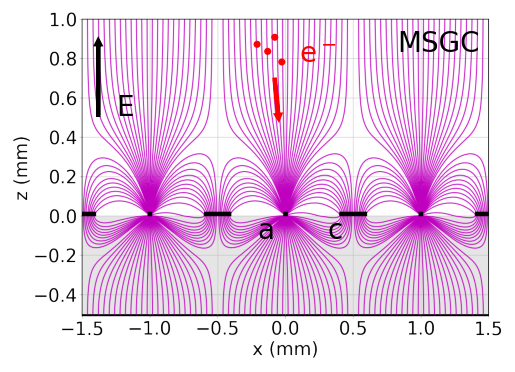} 
    \includegraphics[width=0.49\textwidth]{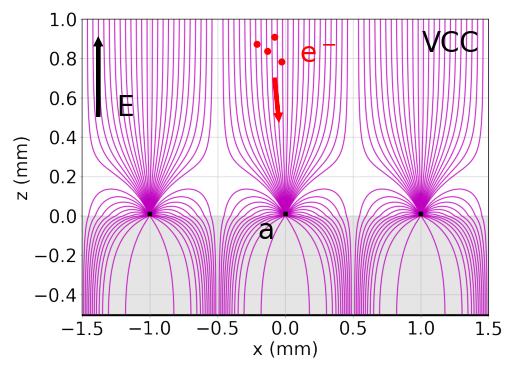}
    \caption{Comparison of the field line patterns of a MSGC (left) and a VCC (right) in liquid xenon calculated with COMSOL\registered\cite{comsol}. The anode strip (a) voltage is 5~kV, the cathode strips (c) are grounded in the MSGC configuration, the backplane is also grounded, and the voltage of the drift electrode (2 mm above strips) is -300~V. The anode-strip width was 8~$\mu$m with a pitch of 1~mm in both calculations, while the width of the cathode strips for the MSGC was set to 400~$\mu$m. The left panel was redrawn and adapted from \cite{Martinez-Lema_2024_microstrips}.}
\label{fig:fields}
\end{figure}

\begin{figure}[h]
    \centering
    \includegraphics[width=0.49\textwidth]{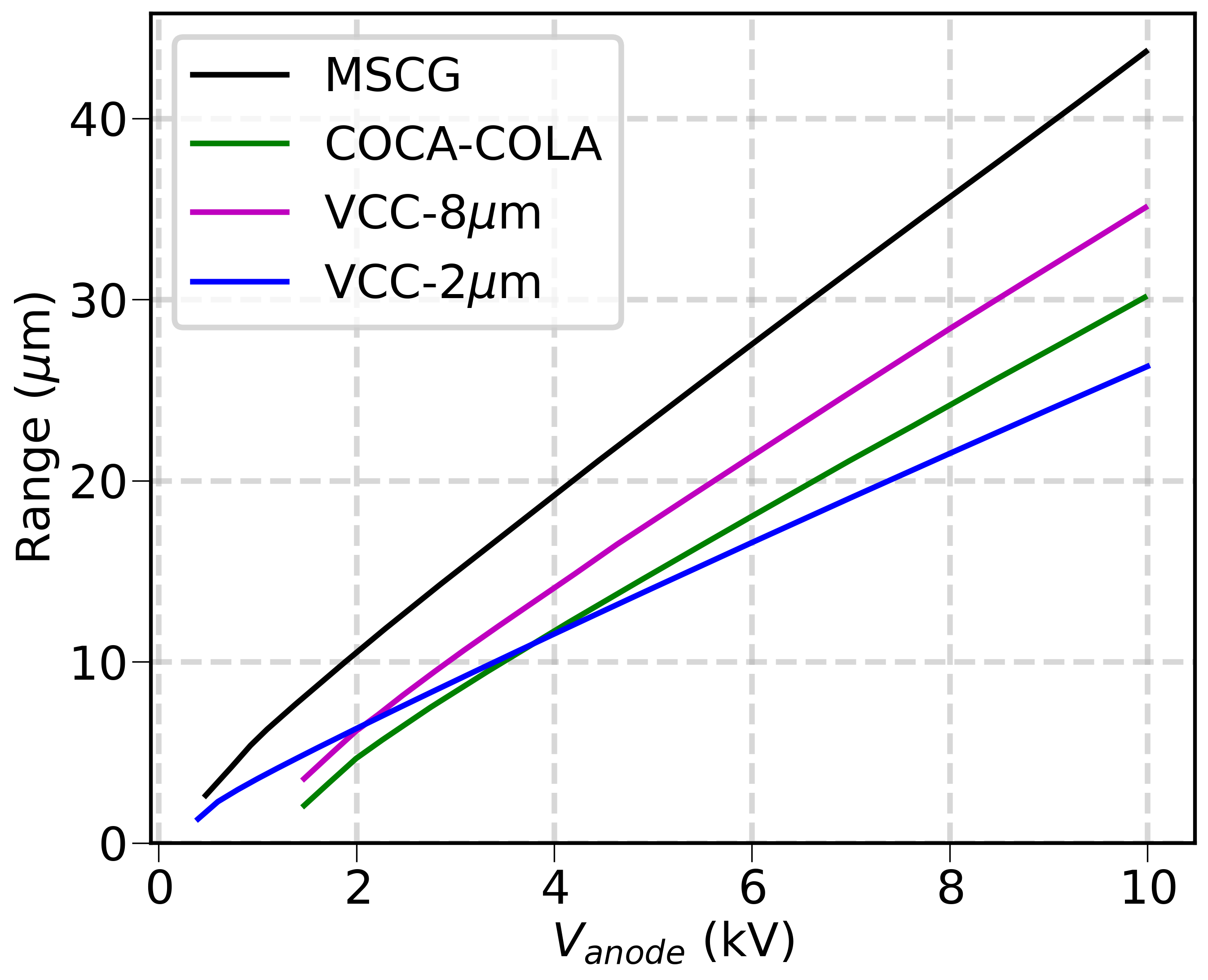} 
    \includegraphics[width=0.49\textwidth]{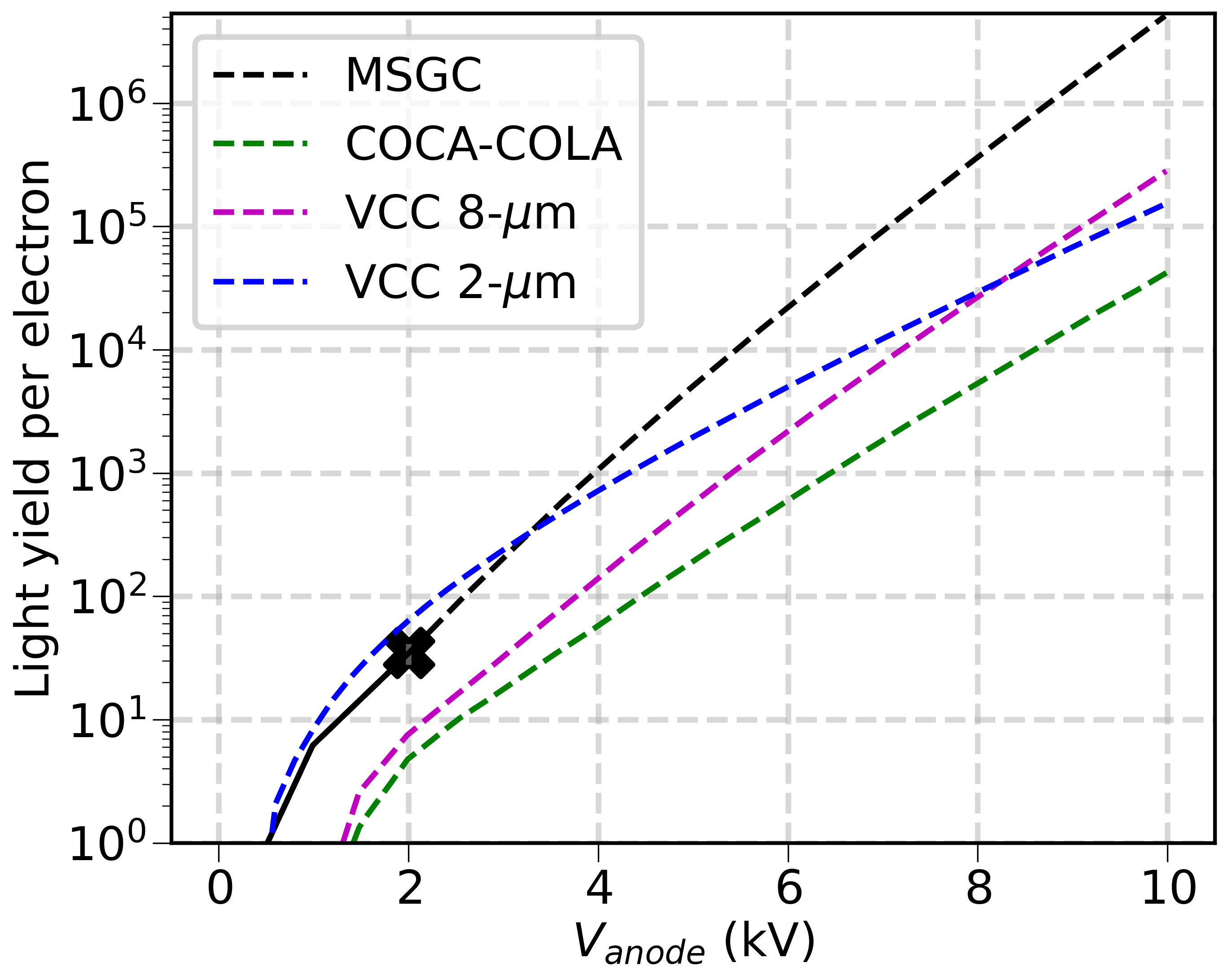} 
    \caption{Comparison of different microstructures in liquid xenon in what concerns secondary light emission. Left: extent of the region along the shortest field line for which the electric field strength is above the electroluminescence threshold measured in \cite{Aprile:2014ELthreshold}. Right: predicted photon yield as a function of the anode voltage. The experimental MSGC result (black cross) was obtained at the highest applicable anode-to-cathode potential. The MSGC, COCA-COLA and VCC-8$\mu$m curves are reproduced from \cite{Martinez-Lema_2024_microstrips} and correspond to plates with 1~mm pitch and 8~$\mu$m-wide anode strips. VCC-2$\mu$m (blue lines) corresponds to the geometry used in this study (2~mm pitch and 2-$\mu$m-wide anode strips). This curve follows the same normalization as the other ones. For further details the reader is referred to \cite{Martinez-Lema_2024_microstrips}.}
\label{fig:prospects}
\end{figure}

Two other MSP configurations were proposed in \cite{Breskin:2022novel}, to permit establishing more intense multiplying fields at the strip vicinity – thus higher expected photon yields.
The two, the coated cathode conductive layer chamber (COCA COLA) \cite{BOUCLIER199174} and the virtual cathode chamber (VCC) \cite{CAPEANS199717} have anode strips deposited on the top surface of the insulating substrate and cathode-strips or cathode-surface at their bottom.
A comparison of the field configuration of the MSGC and VCC designs, obtained with COMSOL\registered, is presented in Fig. \ref{fig:fields}.
These calculations were based on the model suggested in \cite{Aprile:2014ELthreshold} and their established threshold-field values for EL and CM, $\sim$400~kV/cm and $\sim$700~kV/cm, respectively.
The lack of cathode strips on the top surface, combined with the insulator separating the anode strips and the cathode at the bottom, permits applying considerably higher potentials.
COMSOL\registered and model calculations demonstrated the potential superiority of both the COCA-COLA and VCC designs, over the voltage-limited MSGC \cite{Martinez-Lema_2024_microstrips}.
This is illustrated in Fig. \ref{fig:prospects}, where the range for light emission and charge multiplication (left panel) and the light yield per electron (right panel) in LXe are displayed as a function of the voltage applied to the anode strips for these three designs.

In this article, we report on our investigations of a VCC multiplier in liquid xenon. The studied VCC was formed on a semiconductive S8900 glass substrate with 2-$\mu$m wide anode strips.
A maximum initial photon yield of the order of 460~photons/e was reached, decaying in time to $\sim$30~photons/e.
This effect, under investigation, is presumably due to the charging up or polarization of this semiconductive substrate; better stability is expected with other substrate materials that will be the subject of further studies. 

\section{Experimental setup}
\label{sec:setup}

\begin{figure}[h]
\centering
\includegraphics[height=58mm]{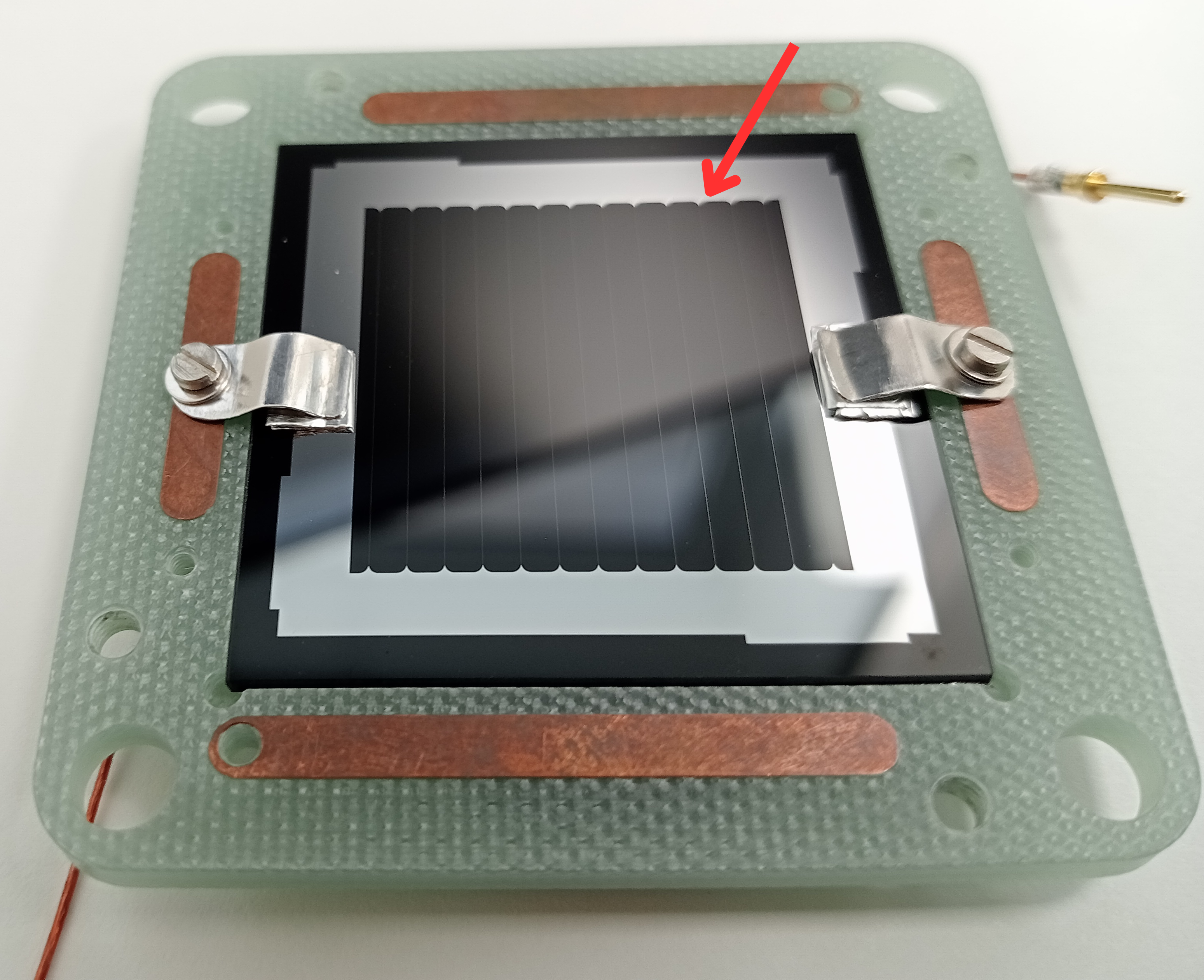} 
\includegraphics[height=58mm]{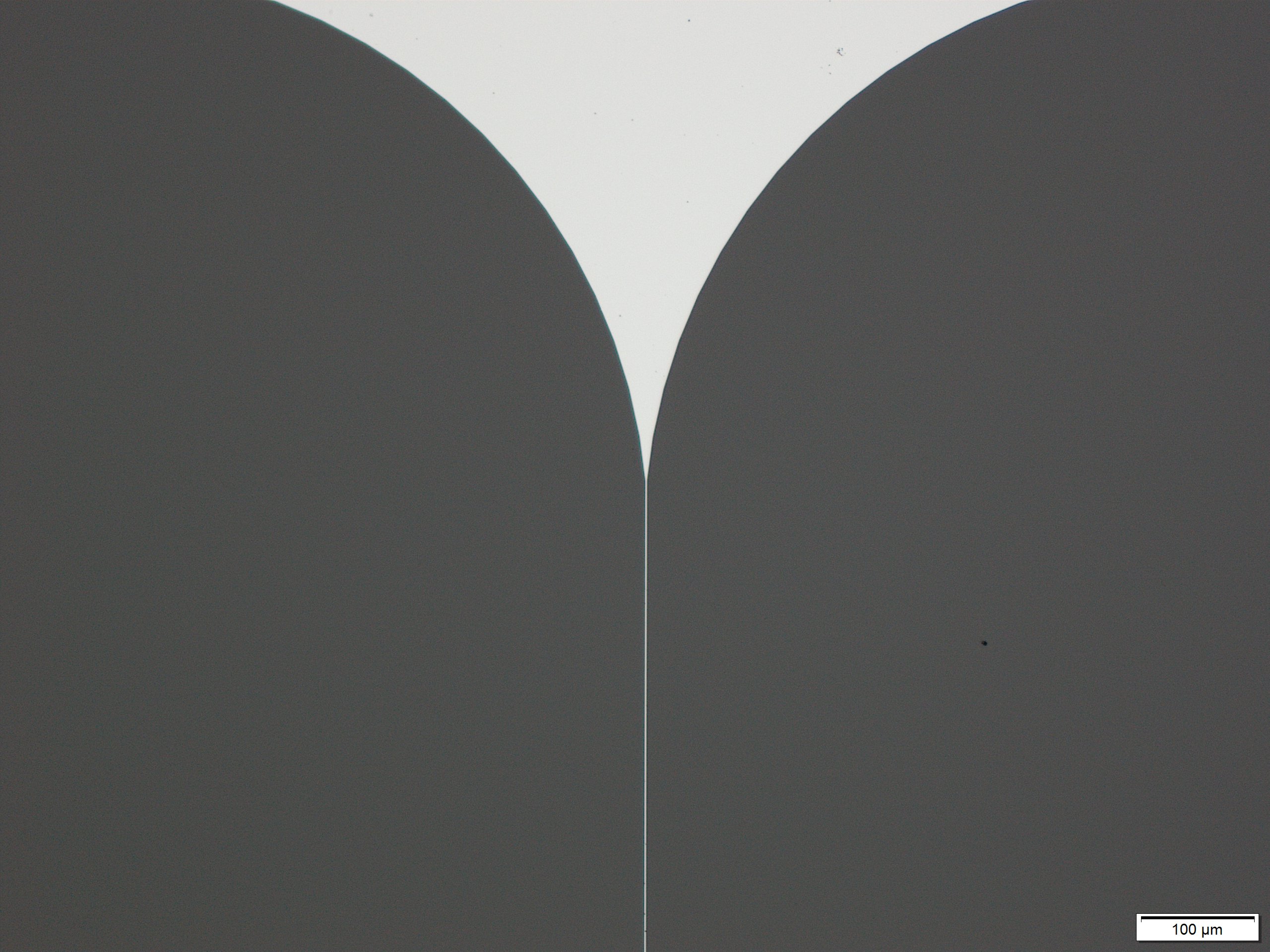} 
\caption{Left: VCC plate mounted on a FR4 support: 13 parallel 2$\mu$m wide strips are connected to a square frame. Two stainless steel clamps provide connection to the external circuit using a multiple-folded aluminium foil to avoid damage to the metal coating. Right: Microphotograph of an anode strip at its junction point with the surrounding frame (indicated by an arrow on the left panel).}
\label{fig:vcc_mount}
\end{figure}

The VCC used in this experiment was produced by IMT\footnote{IMT Masken und Teilungen AG, Greifensee, Switzerland}. Parallel chromium strips of 2~$\mu$m width, all interconnected, were deposited at a pitch of 2~mm on one side of a 1.45~mm-thick 38$\times$38~mm$^2$ glass substrate as shown in the left panel of Fig.~\ref{fig:vcc_mount}. The right panel of Fig.~\ref{fig:vcc_mount} displays a microphotograph of a single anode strip. On the other face of the plate, a continuous metal layer was deposited to serve as the backplane electrode. The thickness of chromium deposition is about 200~nm. The substrate material is a SCHOTT S8900 glass (also known as Pestov glass) commonly used as a substrate for MSGCs \cite{Bouclier:1993, OrtunoPrados:1995}. It is an electron-conducting semiconductor glass with bulk resistivity $\rho=1.1\times 10^{11}\, \Omega \cdot $cm at room temperature \cite{OrtunoPrados:1995, SCHOTT} that results in a total resistance of $\sim 4\times 10^9 \, \Omega\,$ between the strips and the backplane.

The plate was mounted on a FR4 frame, as shown in Fig.~\ref{fig:vcc_mount}, held in place with two stainless steel clamps, which also provided electrical contacts to the strips. To avoid damage to the thin metal coating, an aluminium foil folded several times was placed between the clamps and the plate.

\begin{figure}[h]
\centering
\includegraphics[scale=0.4]{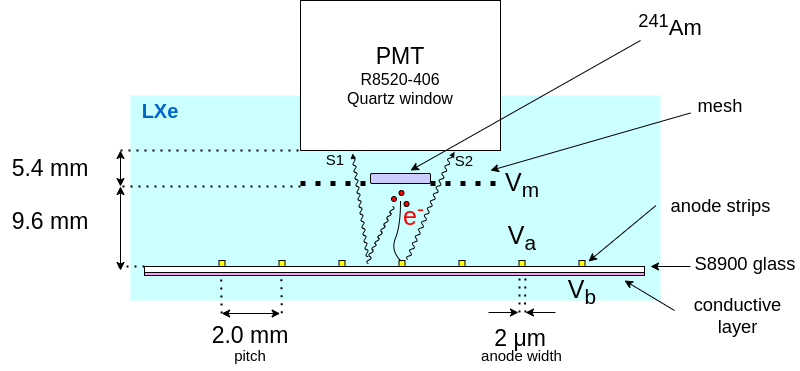} 
\caption{Schematic drawing of the experimental setup. Scintillation photons generated at the interaction site can be reflected from the microstrip plate and reach the PMT, providing the S1 signal. The delayed electroluminescence pulse is generated by the ionization electrons in the vicinity of the strips. The S2 signal results from both photons that reach the PMT directly and those reflected from the strip surface. }
\label{fig:setup}
\end{figure}

A schematic drawing of the whole assembly is shown in the left panel of Fig.~\ref{fig:setup}. A stainless steel mesh electrode on a FR4 frame was mounted 9.6~mm above the VCC plate. The stack was held together using PEEK rods and spacers. A 2-mm-diameter non-spectroscopic $^{241}$Am $\alpha-$source with an activity of $\sim$40~Bq was mounted in the centre of the mesh with the active area facing the VCC. The energy of the emerging $\alpha-$particles was measured to be $(4.7\pm0.2)$~MeV.  

A Hamamatsu R8520-406 \cite{Hamamatsu} 1$\times$1 square inch metal-case photomultiplier with quartz window was mounted 5.4~mm above the mesh. The dimensions were chosen such that the electroluminescence in the liquid (S2), in the vicinity of the strips, could be directly seen by the photomultiplier while a fraction of the primary scintillation photons (S1) emitted at short distance from the source (the range of $\alpha-$particles in liquid xenon is $\sim$40~$\mu$m) could only be observed after reflection from the VCC plate.

The whole assembly was placed inside the cryostat described in detail elsewhere \cite{Erdal:2019Thesis}. Xenon gas was purified by recirculating through a SAES hot getter, model PS3-MT3-R-2, at a rate of 1.2~slpm. The recirculation was kept also during the measurements.

\vspace{0.7cm}

We shall use the following notation for the potentials: $V_a$ for anode strips, $V_b$ for backplane, and $V_m$ for mesh electrode, as shown in Figure \ref{fig:setup}. We shall also use $\Delta V_{ab}=V_a - V_b$ to designate the voltage difference accross the VCC plate. CAEN N471A high voltage power supplies were used for biasing the electrodes with the current limit set at a few nA to protect the VCC from damages due to discharges. This also limited the voltage ramping rate, which was $\sim$200~V/s.

Besides the luminous signal, a charge signal was measured by connecting one of the VCC electrodes to a low-noise charge-sensitive preamplifier ORTEC 142PC.
Typical operation voltage of the PMT was $-$750~V although a significantly lower voltage was used in some measurements in order to reduce electronic pick-up to the charge-sensitive preamplifier. The PMT anode signal was fed directly to a Tektronix 5204B digital oscilloscope.
\section{Results}
\label{sec:results}

\subsection{Light measurements}
\label{sec:light}
\begin{figure}[h]
\centering
\includegraphics[width=\textwidth]{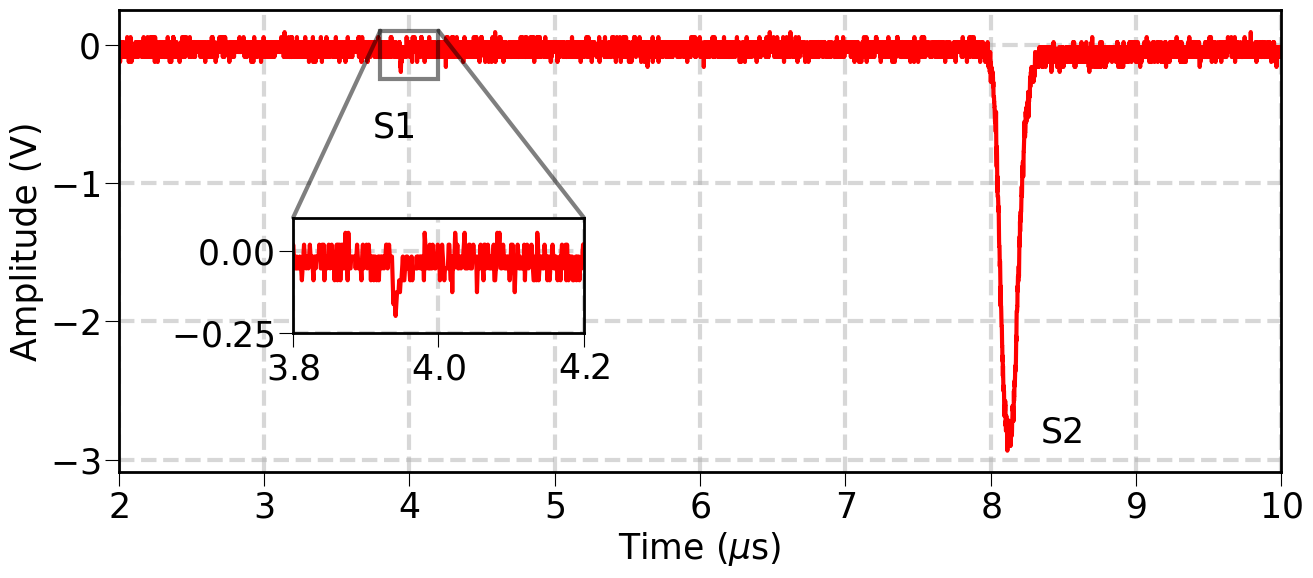} 
\caption{A typical PMT waveform from the VCC in LXe taken at $\dab=5.0$~kV. A primary scintillation pulse (S1, zoomed inset axis) at around $t\approx4~\mu$s is followed by a delayed electroluminescence (S2) signal at $t\approx8~\mu$s. Electrode potentials: $\vmesh=0$, $\vanode=+3.75$~kV, $\vback=-1.25$~kV.}
\label{fig:PMTwaveform}
\end{figure}

An example of the PMT signal is shown in Fig.~\ref{fig:PMTwaveform}. The oscilloscope was triggered by the fast primary scintillation signal (S1) which is followed by the slower but larger pulse produced by secondary scintillation occurring in the vicinity of the VCC anode strips (S2). For each measurement condition, a large number of waveforms were accumulated (typically $\sim$2400) and integrated offline. The PMT was calibrated using a pulsed blue LED, using the method described in \cite{DOSSI2000623}, which allowed for the conversion of the S2 signal area into a number of detected photoelectrons.

\begin{figure}[h]
\centering
\includegraphics[width=0.6\textwidth]{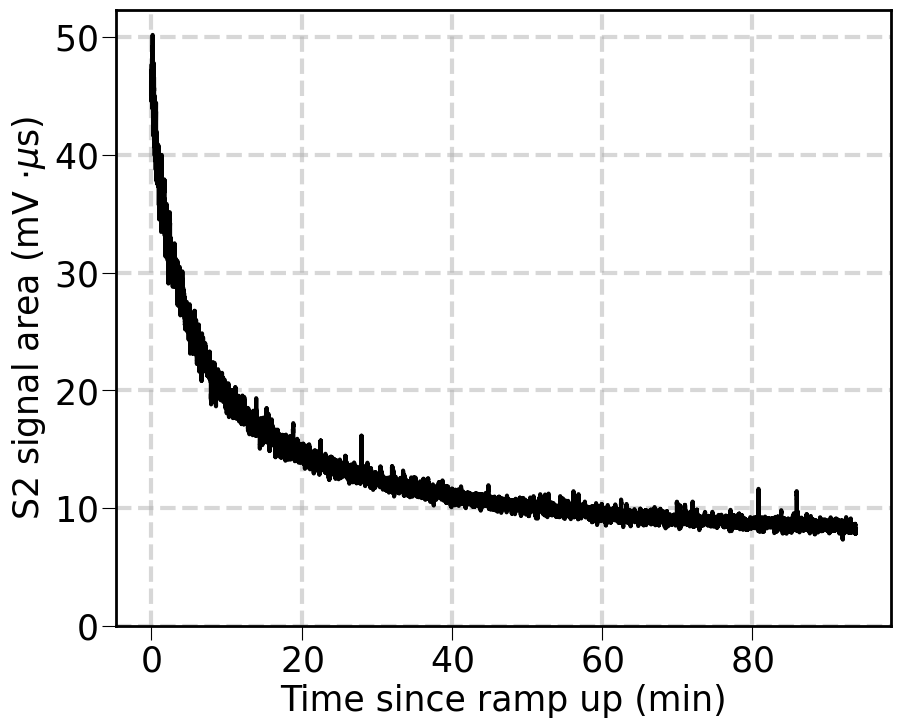} 
\caption{S2 signal area as a function of time after applying voltage across the VCC plate. Initially, $\vmesh=-2$~kV and $\vanode=\vback=0$ were set; then, \vback was ramped up from 0 to -2~kV, keeping \vanode at 0 V. The ramping up time of \vback was about 10~s.}
\label{fig:S2_vs_time}
\end{figure}

During the measurements, we noticed that the S2 signal area decayed significantly over time after setting up the required voltage difference across the VCC plate. This observation is illustrated in Fig.~\ref{fig:S2_vs_time}, which shows the S2 signal area as a function of time at $\vmesh=-2$~kV, $\vanode=0$, and $\vback=-2$~kV. Here, $t=0$ corresponds to the instant when \vback, initially at 0, was turned on.
Given this S2 amplitude decay, discussed in more detail in Section \ref{sec:discussion}, the following protocol has been established when measuring the light yield as a function of \dab: after every change of \dab (in steps of 200~V), a waiting interval of 5~min was allowed after which the waveforms were accumulated during $\sim$1~min.

\begin{figure}[h]
\centering
\includegraphics[width=0.6\textwidth]{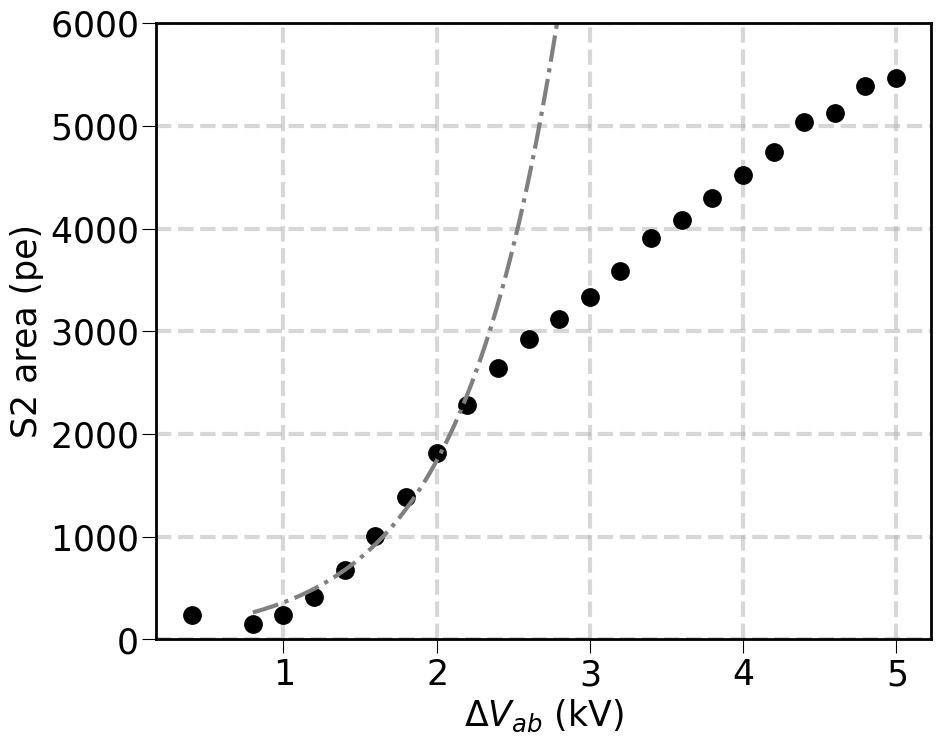} 
\caption{S2 signal area in number of photoelectrons (pe) as a function of the voltage difference between the anode strips and the backplane \dab. Our expectation, based on the EL amplification model (blue line in Fig.~\ref{fig:prospects}) fitted to the data points with $\dab < 2$~kV, is shown as a dashed line. The potentials at the mesh and the anode strips are fixed ($V_m=-2.0$~kV,  $V_a=+3.0$~kV), the backplane potential is varied between $V_b=+3.0$~kV and $V_b=-2.0$~kV. A 5~min interval was allowed after each voltage ramping with a 200~V step.}
\label{fig:LvsDAB}
\end{figure}

The area of S2 signal versus \dab in steps of 200~V, measured as described above is shown in Fig.~\ref{fig:LvsDAB} (black circles). The amplification trend is exponential in the range $\dab = 1.0 - 2.2$~kV. Above $\dab=2.2$~kV we observe an abrupt trend change towards a linear relation between the S2 signal area and \dab. This behaviour, discussed in Section \ref{sec:discussion}, diverges from the EL amplification model (dashed line) used in \cite{Martinez-Lema_2024_microstrips}.

\begin{figure}[h]
\centering
\includegraphics[width=0.49\textwidth]{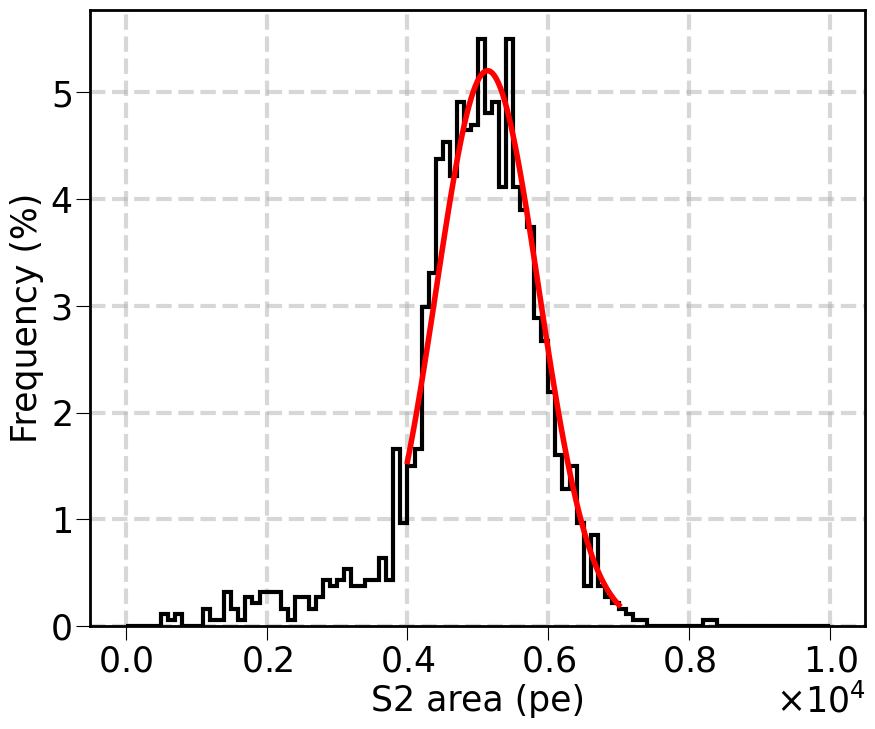} 
\includegraphics[width=0.49\textwidth]{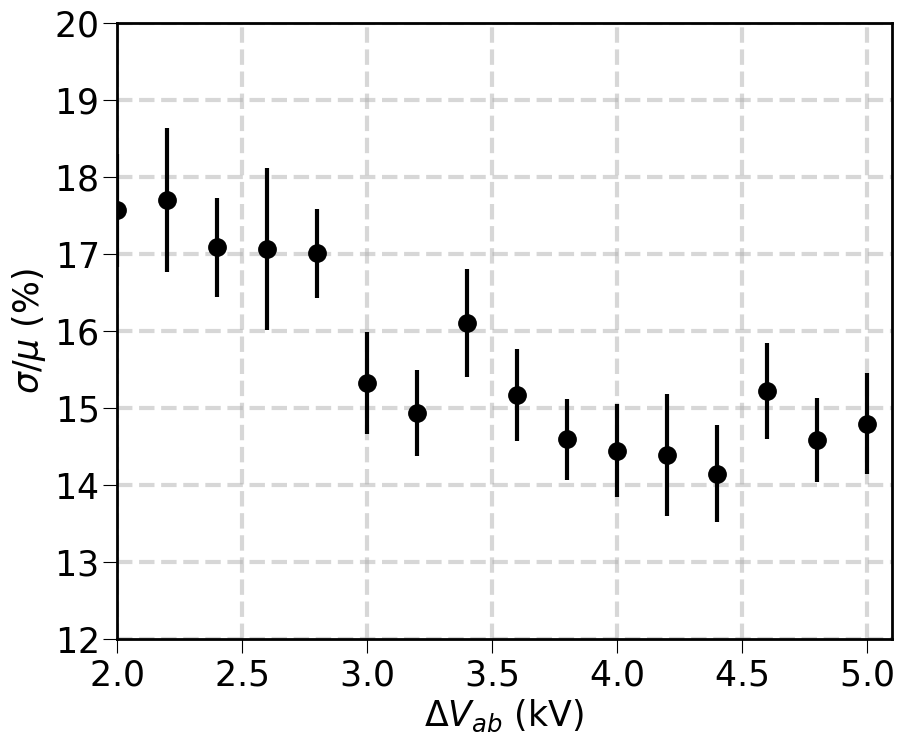} 
\caption{Left: Spectrum of S2 signal areas (black line) obtained at $\dab=4.4$~kV. A fit of a normal distribution to the data is also displayed (red line) yielding an energy resolution of $\sigma/\mu\approx$14 \%. Right: Measured energy resolution as a function of \dab for $\dab \geq 2$~kV. Each spectrum was obtained from 2400 waveforms accumulated during $\sim 1$ min with a 5-minute interval between measurements to allow for the stabilization of the light yield. 
}
\label{fig:energy_resolution}
\end{figure}

The same data were used to assess the energy resolution. A normal distribution with mean $\mu$ and standard deviation $\sigma$ was fitted to the obtained distribution of areas under the S2 signal for each value of \dab. The spectrum for $\dab = 4.4$~kV is shown in the left panel of Fig.~\ref{fig:energy_resolution} (black line) with the gaussian fit superimposed (red line). The dependence of the energy resolution with \dab is shown in the right panel of Fig. \ref{fig:energy_resolution}. The best energy resolution was observed at $\dab = 4.4$~kV, with a value of $\sigma/\mu\approx$14\%.

\subsection{Charge measurements}
\label{sec:charge}

\begin{figure}[h]
\centering
\includegraphics[width=0.9\textwidth]{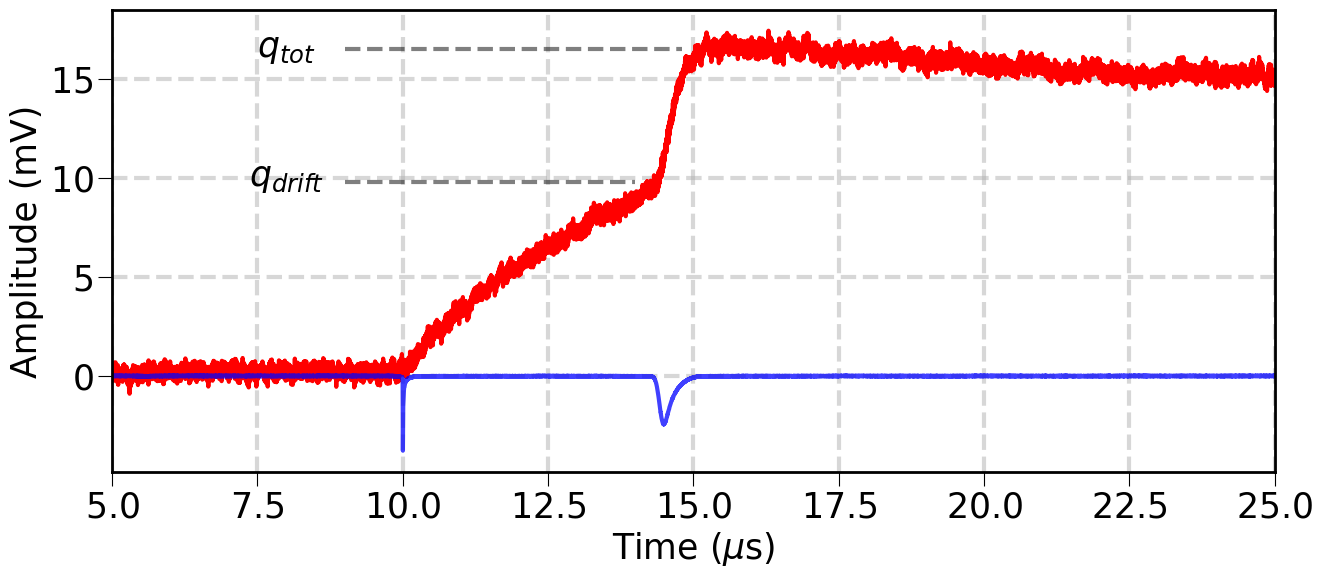} 
\caption{A typical charge-sensitive preamplifier average waveform (red) read from the VCC anode strips in LXe at $\Delta V_{ab}=2.7$~kV. The corresponding PMT average waveform is shown in blue. These data were taken at a lower PMT voltage to minimize electronic pick-up (see text). Electrode potentials: $\vmesh=-2$~kV, $\vback=-2.7$~kV, $\vanode=0$, $\vpmt = -500$~V.}
\label{fig:CSPwaveform}
\end{figure}

Complementary to the light signals, our setup allows for the measurement of charge signals using a charge-sensitive preamplifier. Fig.~\ref{fig:CSPwaveform} shows a typical preamplifier average waveform (red line) connected to the anode strips. Since the charge signal is of the same magnitude as the electronics noise, we obtain the average of all the waveforms in each run (under the same voltages), enhancing the signal-to-noise ratio, which allowed us to perform the analysis. The corresponding PMT average waveform is also shown in blue for reference. Besides electronic noise, preamplifier waveforms were affected by an intense electronic pick-up from the PMT signal, read using the same oscilloscope. This pick-up noise occurred in coincidence with the S1 and S2 pulses detected on the PMT. For this reason, these charge data were taken at a reduced PMT voltage, which allowed us to minimize this effect to a negligible level while still being able to trigger on S1 pulses.

The waveform starts with a flat baseline and begins rising at $t\approx10~\mu$s, in coincidence with the S1 signal from the PMT. The voltage increases as electrons drift toward the anode strips. Deviations from a linear rise are attributed to electron attachment losses in the drift region. A sharp increase in waveform amplitude is observed in coincidence with the S2 signal, indicating the arrival of ionization electrons to the amplification region, where a charge avalanche is induced. After the electrons are collected on the strips, the waveform exhibits a gradual decline, characteristic of the signal decay driven by the RC constant of the preamplifier.

\begin{figure}[h]
\centering
\includegraphics[width=0.49\textwidth]{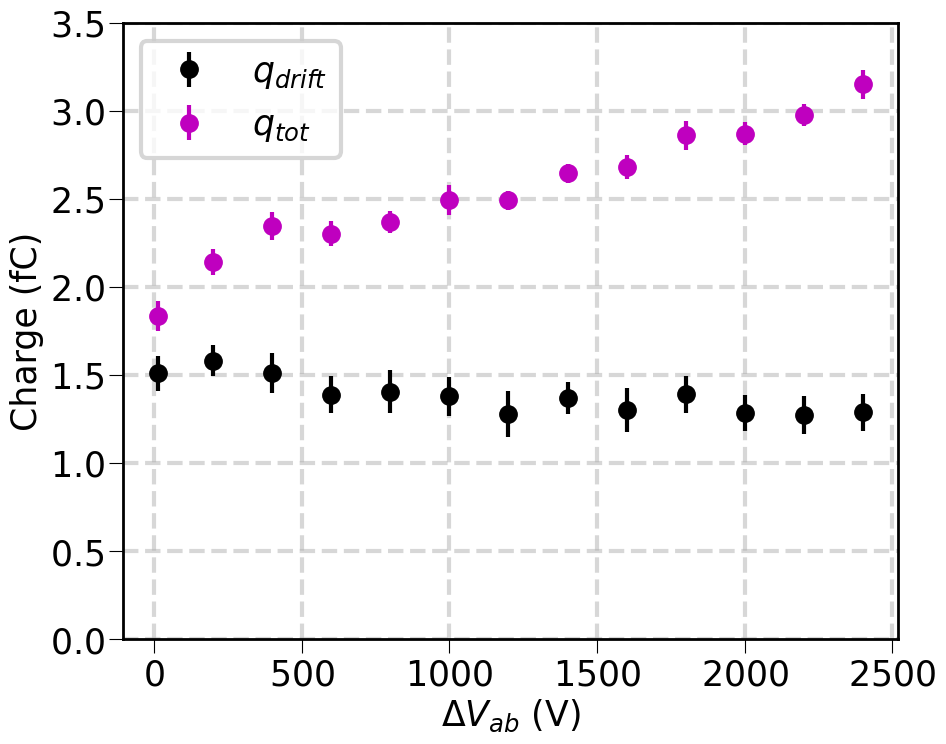}
\includegraphics[width=0.49\textwidth]{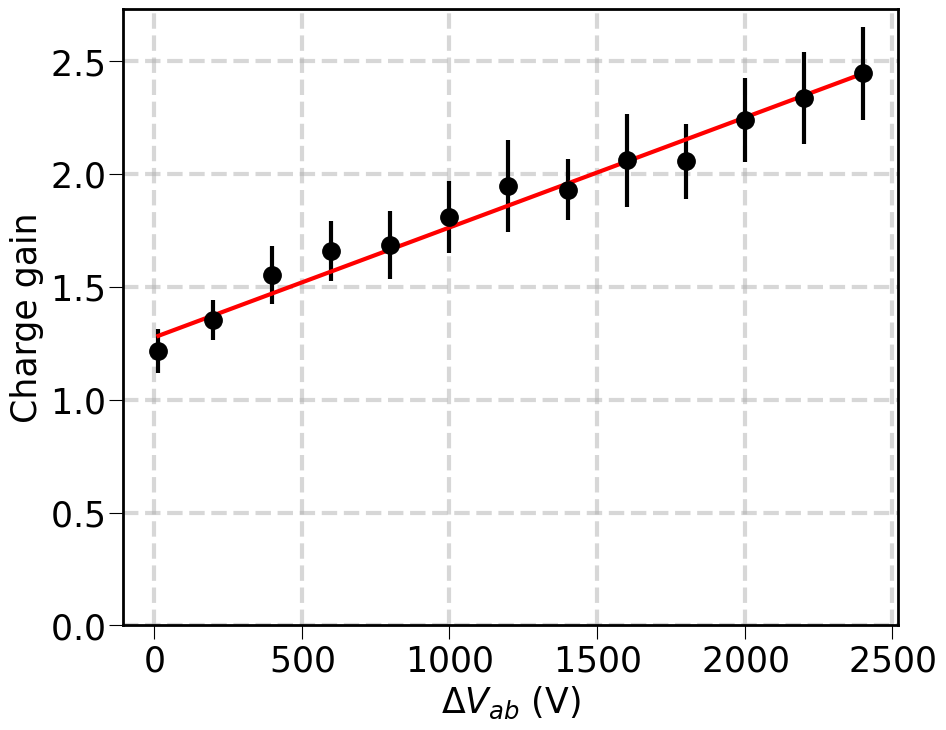}
\caption{Left: estimated drift (black circles) and total (magenta circles) charge collected in the anode strips as a function of the voltage difference between the anode strips and the backplane, \dab. Right: Charge gain (black circles) as a function of \dab with a linear fit to the data superimposed (red line). The potentials at the mesh and the anode strips are fixed ($\vmesh=-2.0$~kV,  $\vanode=0$), and the backplane potential is varied between $\vback=-15$~V and $\vback=-2.5$~kV. A 5~min interval was allowed after each voltage ramping with a 200~V step.}
\label{fig:QvsDAB}
\end{figure}

The charge measured from the anode strips was estimated by measuring the preamplifier waveform voltage before and after the charge multiplication stage, labelled as \qdrift and \qtot in Fig.~\ref{fig:CSPwaveform}. A preamplifier calibration using an external pulser allowed us to convert the waveform amplitude to absolute charge. Due to voltage limitations in our feedthroughs, we could only perform this measurement up to $\dab = 2.5$~kV. The left panel of Fig.~\ref{fig:QvsDAB} shows the estimated \qdrift and \qtot as a function of \dab. A slight reduction of \qdrift with \dab is explained by an increase in the electron recombination due to a weaker drift field near the alpha source when the backplane potential is more negative (i.e. closer to that of the alpha source). This is confirmed by the measured drift time in Fig.~\ref{fig:edrift} (detailed in Section~\ref{sec:discussion}). On the other hand, the total charge \qtot increases monotonically with \dab. The right panel of Fig.~\ref{fig:QvsDAB} displays the estimated charge amplification factor ($\qtot/\qdrift$) as a function of \dab, which is approximately linear. A maximum charge multiplication factor of $\approx$2.5 is observed at $\dab=2.5$~kV. In the studied range, the charge gain increases at a rate of 0.49~kV$^{-1}$. Extrapolating this behaviour to the maximum working voltage achieved, we estimate a charge multiplication factor $\lesssim$5 at \dab~=~5~kV.

\section{Discussion}
\label{sec:discussion}

\begin{figure}
    \centering
    \includegraphics[width=0.7\linewidth]{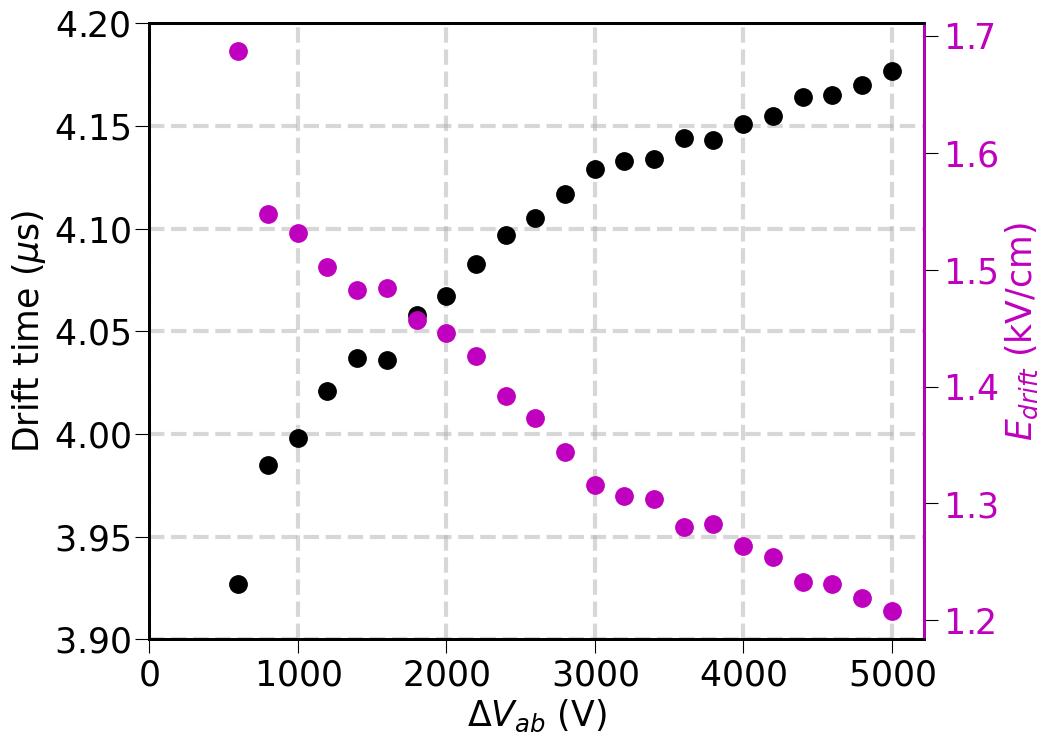}
    \caption{Measured drift time (black) and estimated drift field (magenta) as a function of \dab. The drift field was estimated from the drift time using the experimental data from \cite{Gushchin_1982}.}
    \label{fig:edrift}
\end{figure}

The main advantage of this VCC-like microstrip design is the increased range of operation voltages with respect to to the MSGC-like layout. This feature, given by the lack of cathode strips on the active face, was expected to provide higher light yields as a result of the more intense electric field in the vicinity of the anode strips \cite{Martinez-Lema_2024_microstrips}. Notwithstanding, the electric field is more difficult to determine by computer modelling, as the active area of the strips is much smaller than the backplane area. Therefore, we measured the time difference between S1 and S2 signals, which was translated into a measurement of the drift velocity. The relation between drift velocity and drift field from \cite{Gushchin_1982} was then used to determine the drift field. Fig.~\ref{fig:edrift} shows the measured drift time (black) and the estimated drift field (magenta) as a function of \dab for the data presented in Section~\ref{sec:light}. Finally, from the drift field we could determine the number of electrons escaping recombination along the alpha-particle tracks \cite{APRILE1991119}.

Under the conditions described in Section~\ref{sec:light}, we collect a maximum of 5500 photons at \dab~=~5~kV. The energy of the alpha particles was measured to be ($4.7 \pm 0.2$)~MeV, and we assume a W-value for ionization in liquid xenon of $W_i = (15.6 \pm 0.3)$~eV \cite{PhysRevA.12.1771}. We estimate a drift field in the vicinity of the alpha source for $\vmesh=0$, $\vanode=3$~kV, and $\vback=-2$kV of $\edrift \approx 1.2$ kV/cm, which translates into (3.4 $\pm$ 0.2)\% of electrons escaping recombination \cite{APRILE1991119}. This corresponds to the extraction of $\approx$10$^4$ electrons from an alpha-particle track. This number is 15\% higher than the charge measurement at low \dab presented in Fig.~\ref{fig:QvsDAB}, which could be attributed to electron attachment.
Accounting for the effective solid angle (8.0\%), the reflectivity of the strips and glass estimated from the refractive index using Fresnel's equations%
\footnote{We followed the methodology presented in Appendix A of \cite{Martinez-Lema_2024_etransfer}.}
($\approx$10\%), the transparency of the mesh holding the alpha source (81\%), and the QE of the PMT photocathode (28\%) \cite{Hamamatsu}, we estimate a light yield of (27.0 $\pm$ 3.1) photons per drifting electron after detector stabilization (Fig.~\ref{fig:S2_vs_time}).
This number is comparable to that obtained in liquid xenon with the MSGC microstrip plate \cite{Martinez-Lema_2024_microstrips} and with thin wires recently \cite{Qi:2023, Tönnies_2024} at similar voltages.

The measured light yield is significantly lower than expected based on our computations (Fig~\ref{fig:prospects}). However, the data were affected by a rapid degradation of the photon yield in S2 after the voltage application to the VCC (see Fig~\ref{fig:S2_vs_time}). Assuming that the cause of this degradation is eliminated, the potential light yield might be much higher.
For example, at $\dab=5.0$~kV, we detect a maximum of $9.3 \cdot 10^4$ photoelectrons before gain degradation. This number translates into a light yield of $\approx$460 photons per drifting electron, which is still about four-fold lower than our prediction in Fig.~\ref{fig:prospects}. We hypothesize that the reasons for this discrepancy may be related to an incomplete modelling of the setup in COMSOL\registered and inaccuracies in the electroluminescence and charge multiplication model we use.

The gain decrease (Fig.~\ref{fig:S2_vs_time}) is the main issue at the present point requiring further investigation. Likely related to the properties of the glass substrate, one may suggest two possible reasons for such behaviour: charging up of the glass surface near the anode strips and a slow polarization time of the glass upon voltage application. It is worth noting that this behaviour was not observed in the MSGC detector previously tested by us \cite{Martinez-Lema_2024_microstrips} where the anode and the cathode strips were deposited on the same side of a D263 Schott glass substrate.

\begin{figure}
    \centering
    \includegraphics[width=0.6\linewidth]{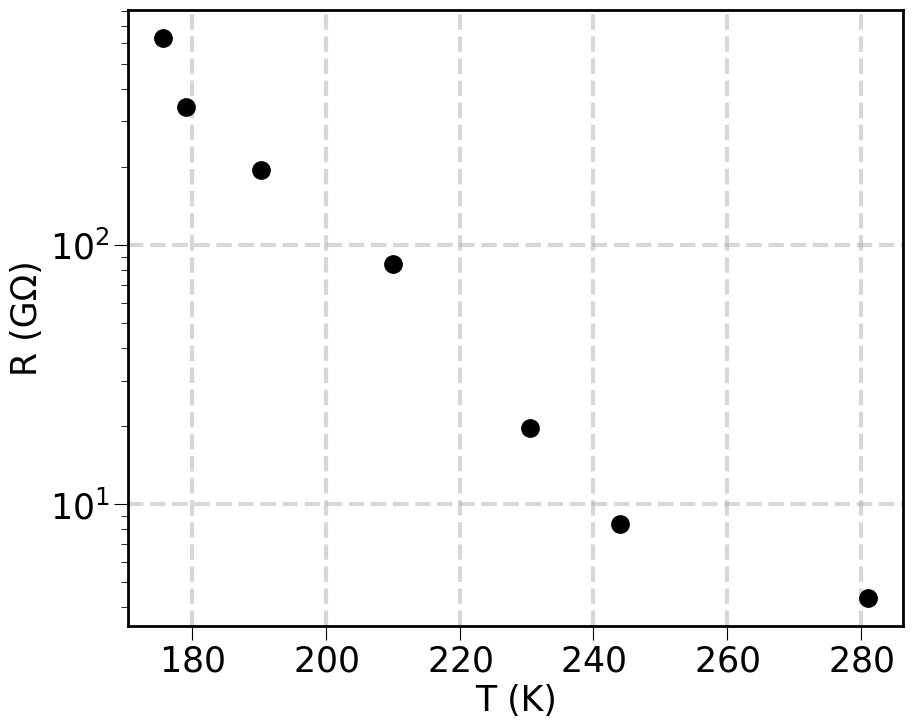}
    \caption{Measured resistance between anode strips and backplane as a function of the temperature in the cryostat. The temperature was estimated as the average between the measurement at the bottom of the cryostat and the cooling ring at the top.}
    \label{fig:r_vs_t}
\end{figure}

The charging up may occur as a result of production of positive ions near the anode strips due to charge multiplication. The field configuration (Fig.~\ref{fig:fields}) and the low conductivity of the glass favour this effect. At room temperature, the bulk resistivity of SCHOTT S8900 glass is $\rho\sim 1.1\times 10^{11} ~\Omega\cdot$cm \cite{Bouclier:1993, OrtunoPrados:1995}. For our VCC plate, we monitored the resistance (R) between the anode strips and the backplane during the cooling down process by measuring the electric current under applied voltage (Fig. \ref{fig:r_vs_t}). At room temperature, we measured $R = 4\times 10^9~\Omega$; at $T\approx$170~K we found $R >6\cdot 10^{11} \Omega$. In terms of bulk resistivity this would be equivalent to a value of  $\rho\geq 1.5\times 10^{13}~\Omega\cdot$cm. The measured trend is consistent with the exponential behaviour observed in similar materials  \cite{Pestov:1994, Blaha:1998, WANG2010151, Roy_2019}.

Although the charge multiplication gain was $\lesssim$5 and the activity of our $^{241}$Am source was only 40~Bq, charging up of the substrate is still a possibility. Positive ions produced in the avalanche can concentrate in the substrate in between strips, creating a positive surface charge and distorting the local electric field (see Fig.~\ref{fig:fields}). These ions would then be neutralized by the current flowing through the substrate. However, if the substrate conductivity is not sufficient to neutralize the ions promptly, there will be a net concentration of positive charges on the glass surface between the anode strips. This would result in deviation of some part of the drifting electrons from the strips and in a reduction of the observable light and charge gains. These electrons would recombine with the ions on the glass surface, eventually reaching an equilibrium in a stable, but distorted, field configuration.

Gain variations in time have been widely reported for MSGC in gas. In some cases, the gain depended on the irradiation rate, but in the others the gain was only a function of time after applying the bias voltage independently of the irradiation rate. Both gain drop and gain increase over time have been observed; other more complex behaviours have also been noticed. For further details, we refer for example to \cite{BOUCLIER199174, BOUCLIER:1992, Bateman:1992, Bouclier:1993, Pestov:1994, RD28:1996} and the references therein.

Gain changes after voltage application and at a minimal exposure to radiation (with no irradiation, virtually) have been noticed in most of the MSGC on dielectric supports. As reported in \cite{Bouclier:1993, Pestov:1994}, for MSGCs on substrates made of ionic glasses with $\rho\sim 10^{15}~\Omega\cdot$cm, a gain drop of 15\% was observed on a time scale of about 1~hour while for an electronically conductive glass (similar to the one used here) with $\rho\sim 10^{9}-10^{12}~\Omega\cdot$cm, no changes were detected (all measurements in gas at room temperature). A possible mechanism for such a variation is the polarization of the high-resistivity material, which behaves as a dielectric upon application of the voltage across it.

\begin{figure}
    \centering
    \includegraphics[width=0.7\textwidth]{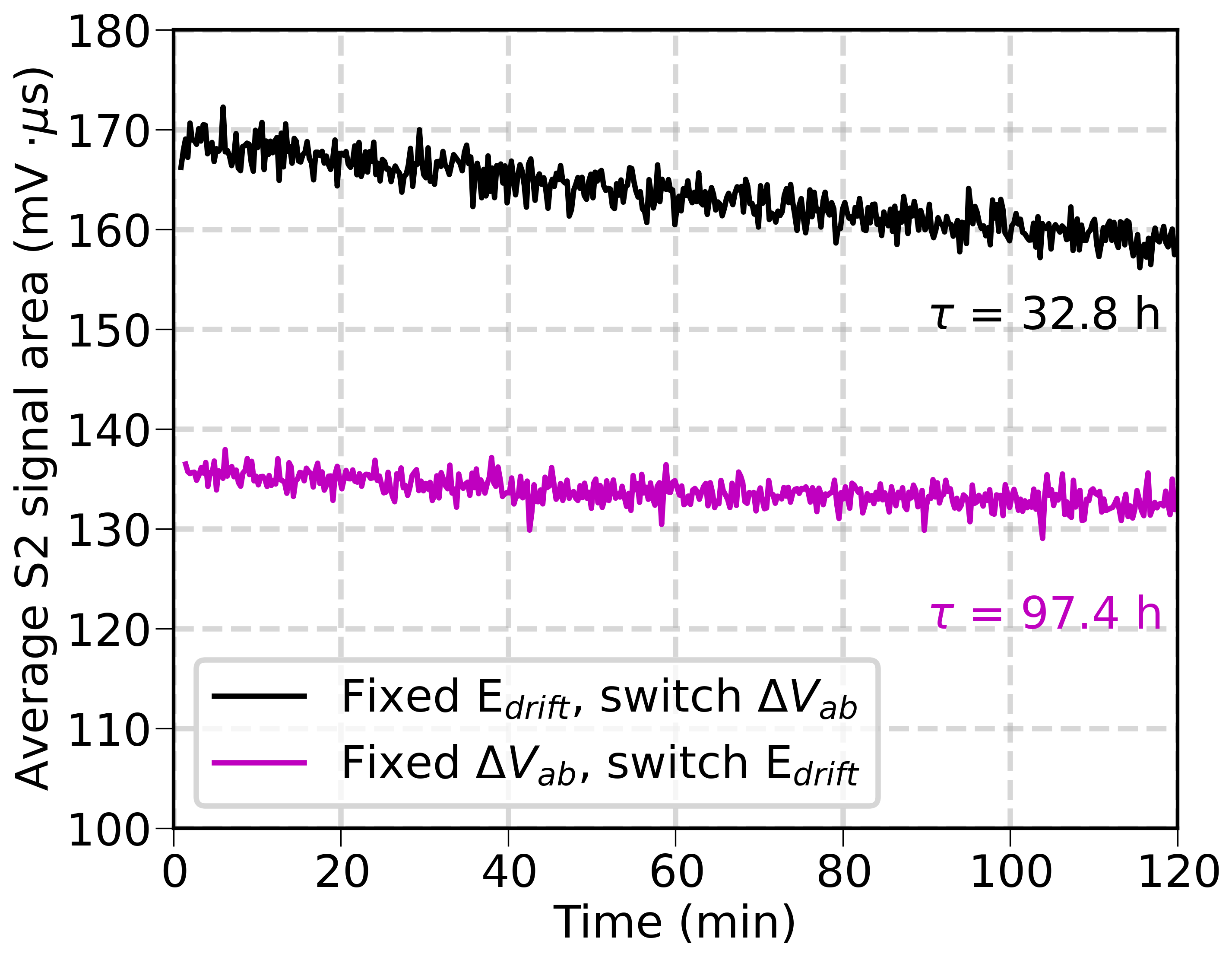}
    \caption{S2 signal area as a function of time of the VCC in xenon gas at room temperature after ramping \dab for a fixed \vmesh (black) and after ramping \vmesh for a fixed \dab (magenta). An exponential function is fitted to each dataset, from which a decay constant ($\tau$) is extracted.}
    \label{fig:s2_vs_time_gas}
\end{figure}

As an attempt to shed some light on the mechanism of the S2 reduction, we measured the S2 light yield in xenon gas at room temperature. In our configuration (Fig.~\ref{fig:setup}), the range of alpha particles at 2 bar is longer than the gap between the source and the strips; therefore, alpha-particle emission over a broad solid angle resulted in a broad energy spectrum.
Nonetheless, this feature is not time-dependent. We observed a much smaller time-dependent decay than in liquid. Fig.~\ref{fig:s2_vs_time_gas} depicts the average of the S2 signal area as a function of time after ramping up \dab (the voltage across the VCC) with a fixed \vmesh (left) and after ramping \vmesh with a fixed  \dab (right). For the former, a decay time of the order of 30 h is observed, while for the latter, the decay time is approximately 3 times longer. This behaviour is compatible with that reported in \cite{Bouclier:1993}. The ratio between these two constants suggests that charging up effects (which should be stronger in gas due to the higher ionization-electron yield from the alpha-particle track and higher charge gain) play a secondary role, and that polarization of the substrate is a more fitting explanation.

\section{Conclusions}
\label{sec:conclusions}

We have validated the novel approach of electroluminescence amplification in liquid xenon with a VCC (Virtual Cathode Chamber) strip plate. The initial light yield, upon potentials application, was $\sim$460 photons per electron. However, the photon yield decayed by a factor of $\sim$17 within $\sim$3~h to a stable value of $(27.0 \pm 3.1)$ photons per drifting electron. This suggests that high photon yields could be achieved, provided the cause of this decay is understood and eliminated. 

The electroluminescence of liquid xenon was accompanied by charge multiplication in the vicinity of the anode strips. Extrapolating the charge gain measurements at lower potentials, we estimated a charge gain factor $\lesssim$5 at $\dab = 5$~kV.

Taking into account the known experience of operation of microstrip gaseous chambers (MSGC), two possible reasons for the degradation of the photon yield could be the charge accumulation on the glass surface and the glass polarization. Further research is necessary to identify the reasons for the observed gain drop and to improve the electroluminescence yield. In particular, the investigation of other substrate materials have the potential of improving the performance of this device, as substrates with lower resistivity are likely to perform better.
VUV-transparent substrates (e.g. quartz, MgF2, BaF2 etc.) should also be investigated to enable detection of primary scintillation and xenon electroluminescence induced near the strips e.g. in detector configurations proposed in \cite{Breskin:2022novel}. We expect VCC-based single-phase noble-liquid detectors to become potential tools in dark-matter searches, neutrino physics and other fields.

%%%%%%%%%%%%%%%%%%%%%%%%%%%%%%%%
% APPENDICES
%%%%%%%%%%%%%%%%%%%%%%%%%%%%%%%%
\appendix

%%%%%%%%%%%%%%%%%%%%%%%%%%%%%%%%
% ACKNOWLEDGMENTS
%%%%%%%%%%%%%%%%%%%%%%%%%%%%%%%%
\acknowledgments
This work was performed at the Detector Physics Laboratory of the Weizmann Institute of Science (WIS), in the context of the DARWIN dark-matter and CERN-RD51 collaborations. It was supported in part by Fundação para a Ciência e Tecnologia through projects CERN/FIS-INS/0013/2021, 2024.00269.CERN, CERN-RD51 Common Fund Grant, the CERN GDD group, and The Krenter-Perinot Center for Experimental High Energy Physics (WIS). G.M.L. acknowledges the personal support of Dr. L. Arazi of Ben Gurion University.
V.C. acknowledges the personal support of the WIS Visiting Professor Program.

%%%%%%%%%%%%%%%%%%%%%%%%%%%%%%%%
% BIBLIOGRAPHY
%%%%%%%%%%%%%%%%%%%%%%%%%%%%%%%%
\bibliographystyle{JHEP}
\bibliography{bibliography}

\end{document}